\documentclass[12pt,a4paper]{article}
\usepackage{amsmath}
\usepackage{graphicx}
\usepackage{times}
\begin{document}
\begin{titlepage}
\title{Experimental signatures of hadron asymptotics at the LHC }
\author{ S.M. Troshin, N.E. Tyurin\\[1ex]
\small  \it SRC IHEP of NRC ``Kurchatov Institute''\\
\small  \it Protvino, 142281, Russian Federation}
\normalsize
\date{}
\maketitle

\begin{abstract}
We  discuss  experimental signatures relevant for the LHC domain of energies allowing to discriminate possible asymptotic modes of hadron interactions. 
\end{abstract}
\end{titlepage}
\setcounter{page}{2}
\section*{Introduction}
The studies of global geometrical properties of hadron interactions    represent an important step \cite{blg} toward to  understanding of hadron dynamics related to the  development of QCD in the nonperturbative region, in particular, aimed to the soft hadron interactions description. Those properties are encoded in the impact--parameter dependencies of the total, elastic and inelastic overlap functions. These functions depend on two  variables---energy and impact parameter--- and their knowledge provides therefore more information on hadron interactions compared to the integrated over impact parameter observables. 

Such observables as cross-sections and average impact parameters (the slope of the diffraction cone is determined by the total average impact parameter squared) are the important global though incomplete and indirect characteristics of the geometry of hadron interactions  but their measurements   can be  performed experimentally. 
However, despite its experimental accessibility,
integrated observables provide a limited knowledge of the hadron interaction domain geometry in the impact parameter representation. Moreover, the integrated observable with the same energy dependence can correspond to the  different physical situations, e.g., the limiting value of the ratio $\sigma_{el}(s)/\sigma_{tot}(s)\to 1/2$ at $s\to\infty$ does not unambiguously implies that the hadron interaction corresponds to the black-disk  model.  It has been shown in \cite{drw}. 

The above conclusion is most relevant when the consideration is narrowed and limited by the analysis of the energy dependence of cross--sections only  and sometimes of the mean multiplicity ( cf. \cite{stod} and the references therein). Evidently, there is only one-way road from the impact parameter picture to the predictions for integrated observables and moving in the opposite direction is an illegitimate and logically inconsistent operation.

 For a long period of time the hadron elastic interactions  have been considered as consistent with the picture denoted by the acronym BEL, i.e. it supposes  that the protons' interaction region becomes Blacker,  Edgier and Larger \cite{valin1} with increasing energy. 
 Such energy evolution can explain  that the energy dependence of the diffraction cone shrinkage is slower than the  growth of the total cross-section.  Asymptotically they should have similar energy dependence (cf. \cite{fag} and references therein).  This means that the cone shrinkage should be accelerated when the energy values  are close  to the asymptotic region \cite{rysk}.
 
 On the base of  rational (i.e. non exponential) unitarization scheme, it was noted in \cite{edn,npwe} that the inelastic overlap function with the increasing energy could acquire  a peripheral form in  the impact parameter representation. Such form has been interpreted then as a manifestation of emerging interaction transparency in the central hadron collisions. Later on this interpretation has been generalized  in papers \cite{bd1,bd2,bd3} where this phenomenon  has been treated  as antishadowing or reflective scattering when the elastic interactions dominate under  central collisions.    The analysis \cite{alkin} of the elastic scattering data obtained by the TOTEM  at $\sqrt{s}=7$ TeV indicated  on the transition   to the particularly  new scattering mode  \cite{bd1,bd2,bd3,degr,degr1,anis}  currently named as antishadowing, reflective or resonant mode. There is no commonly accepted name for this mode but its existence and gradual transition  to it is under diskussion now in many papers (cf. e.g. the above references) with proposed various interpretations (\cite{edg,alb,arr}).
The peripheral impact parameter dependence  of the inelastic overlap function  with maximum at non-zero impact parameter value (i.e. with depletion of inelastic probability at small impact parameters) was  depicted and diskussed already in \cite{npwe} for the energy value of $\sqrt{s}=3$ TeV. 
 Moreover, as it was pointed out recently in \cite{epja}, there are now two independent impact parameter analyses of the  TOTEM data \cite{alkin,tot8} at the LHC energies $\sqrt{s}=7$ TeV and $8$ TeV indicating on the transition to the reflective scattering mode \cite{bd3}.

Despite the impact parameter analysis indicated existence of the reflective scattering mode at the LHC, this conclusion is not yet commonly accepted. One of the reason is rational and based on the fact that an impact parameter analysis invokes some additional assumptions though quite natural ones. The question: which limit for the scattering amplitude (if any) is saturated at asymptotics, the black disk limit or the unitarity limit is considered to be still unanswered. 

It seems that to provide a more convincing answer  one needs to present some additional arguments based on the directly detectable experimental effects. Assuming the unitarity limit saturation, two more questions have to be answered.  What is the value of the collision energy where the black disk limit for the scattering amplitude being crossed? Does this crossing occur at the LHC? 

In this  note we are going to enlist several  effects that could be experimentally verified and helpful in getting a more affirmative answers to the questions stated above.  In other words, 
is the elastic scattering at the LHC absorptive or geometric one \cite{prdwe13} ?

\section{Experimental signatures of  absorptive  mode}
Since our consideration is qualitative and the high energy experimental data are in favor of the pure 
imaginary amplitude we will suppose in what follows that the real part of the elastic scattering amplitude is small and can be neglected\footnote{It is not quite correct assumption in view of dispersion relation and it can be corrected by the restoration of the real part of the scattering amplitude with the scenario described in \cite{real}.}.

Then replacing $f\to if$  where $f(s,b)$ is the elastic scattering amplitude,  one obtains that the unitarity relation allows to express the inelastic overlap function $h_{inel}(s,b)$
in the form
\begin{equation}\label{un1}
h_{inel}(s,b)=f(s,b)[1-f(s,b)].
\end{equation} 
Note, that $P(s,b)\equiv 4h_{inel}(s,b)$ is the probability of the inelastic hadron collision with the values of energy $s$ and impact parameter $b$.
Unitarity (in this particular normalization) allows variation of $f$ in the interval
$0\leq f \leq 1$. The absorptive scattering mode assumes that the region of elastic scattering amplitude  variation is reduced to $0\leq f \leq 1/2$. The value $f=1/2$ corresponds to the complete absorption of the initial state, i.e. respective elastic scattering element of the scattering matrix is zero, $S=0$  ($S=1-2f$). It is commonly accepted that this limit (black disk limit) has been reached at
small $b\sim 0$ values of the impact parameters at the LHC energies. 

The opinions on further energy evolution are divergent. The first is that this value $S=0$  will be frozen at $b=0$ and further energy increase will make distribution of $f(s,b)$ over impact parameter  wider keeping the same maximal value $1/2$. Another people believe that the energy increase will lead in addition to crossing the value of $1/2$. The two curves corresponding the two above cases are depicted at Fig. 1.
 \begin{figure}[hbt]
	\begin{center}
		\resizebox{12cm}{!}{\includegraphics*{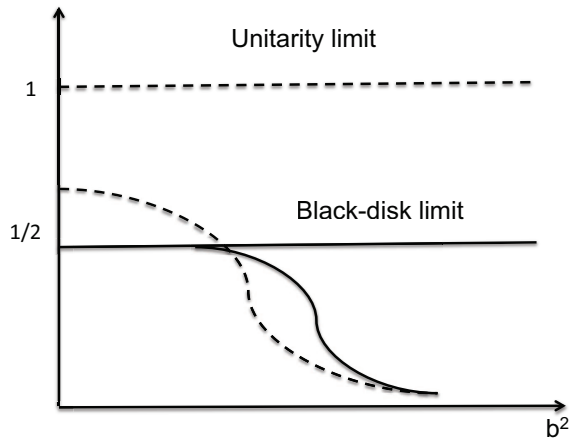}}
	\end{center}
	\vspace{-2cm}
	\caption[ch2]{\small Impact-parameter dependence of the amplitude $f(s,b)$ in the two cases of absorptive (solid line) and reflective (dashed line) scattering modes at the LHC.}
\end{figure}

The reason for coexistence of the two views is that a possible excess of the black disk limit at the LHC energy $\sqrt{s}=7$ TeV is rather small.
The analysis  \cite{alkin} has shown that
$f(s,b)$ is greater than the black-disk limit of $1/2$ at $\sqrt{s}=7$ TeV with the relative excess $\alpha$ ($f(s,b)=1/2[1+\alpha(s,b)]$) and  value of $\alpha$ is  about $0.08$  at $b=0$.  

The measurements of the diffraction cone slope is crucial in this situation. Indeed, this observable
\[
B(s)=\frac{d}{dt}\ln \frac{d\sigma}{dt}|_{t=0}
\]
is proportional to the mean value the impact parameter squared $b^2$, 
\[
\langle b^2\rangle=\frac{\int_0^\infty b^3dbf(s,b)}{\int_0^\infty bdbf(s,b)}.
\]
The Fig. 1 assumes that the forward slope parameter $B(s)$  should start to increase with the energy faster in case of the absorptive scattering mode since the black--disk limit is already reached.
Thus, the asymptotical black--disk limit saturation can be detected through a changing energy dependence of the observable $B(s)$ 
\begin{figure}[hbt]
\begin{center}
\hspace{0.5cm}
\resizebox{10cm}{!}{\includegraphics*{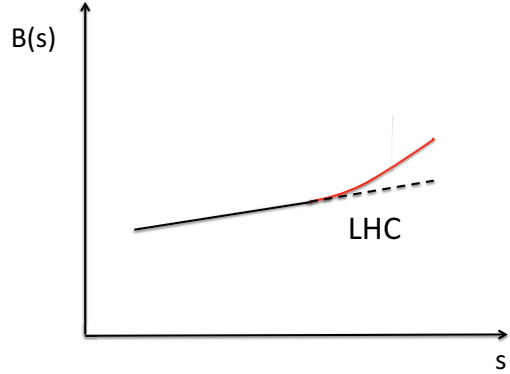}}
\end{center}
\vspace{-2cm}
\caption[ch2]{\small Schematic energy variation of the slope parameter $B(s)$  at the LHC in the two cases of absorptive (solid line) and reflective (dashed line) scattering modes realisations at $s\to \infty$. }
\end{figure} at the LHC energies, namely, $B(s)$ should start to increase faster than it does at lower energies in case of absorptive scattering mode. This increase should be the same as an increase of the total cross-section, i.e. one should expect that the ratio $\sigma_{tot}/B(s)$ would become a constant.
No change of its energy evolution is expected in case of reflective scattering mode. The above conclusions are based on the relation
\begin{equation}\label{stot}
f(s,b)=\frac{1}{4\pi}\frac{d\sigma_{tot}}{db^2}
\end{equation}
being valid for  pure imaginary elastic scattering amplitude $f(s,b)$ and the schematic energy dependence of the slope parameter $B(s)$ is depicted at Fig. 2.
Absorptive mode leads to appearance of the secondary dips in the differential cross-section of elastic $pp$--scattering at the LHC energies and
Fig. 3 
\begin{figure}[hbt]
\begin{center}
\hspace{0.5cm}
\resizebox{10cm}{!}{\includegraphics*{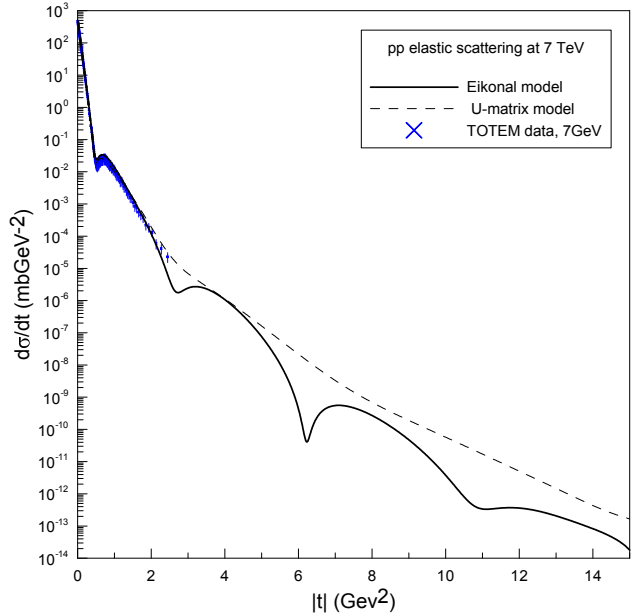}}
\end{center}
\vspace{-3.5cm}
\caption[ch2]{\small Description \cite{martyn} of differential cross--section of $pp$-scattering at $\sqrt{s}=7$ TeV in absorptive (solid line) and reflective (dashed line) forms of unitarization.}
\end{figure}
 is an illustration of this  conclusion. The absorptive  models which are based on the standard eikonal unitarization predict lower values for the differential cross-section in $pp$-scattering at $\sqrt{s}=7$ TeV and appearance of the secondary bumps and dips  at large values of $-t$.

The difference in the behavior of $d\sigma/dt$ in the deep-elastic scattering region results from difference of  forms of the scattering amplitudes in the impact parameter representation. 

Thus, one should emphasize that the typical experimental signatures of the absorptive scattering mode would be change (faster increase) in the energy evolution of the slope parameter $B(s)$ and in presence of the secondary dips and bumps in $d\sigma/dt$ at the LHC energies.

\section{Experimental signatures of reflective  mode}
In this section we consider typical experimental signatures of the reflective scattering mode.
The distinctive feature of the transition to the reflective scattering mode is the developing peripheral form of the inelastic overlap function. This form starts to appear when $f\geq 1/2$.
Schematically energy evolution of the inelastic overlap function is depicted in Fig. 4.
  \begin{figure}[h]
	\begin{center}
		\resizebox{8cm}{!}{\includegraphics*{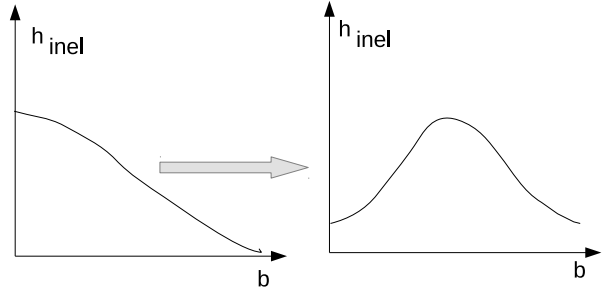}}
	\end{center}
	\caption[ch1]{Energy evolution of $b$--dependence of the inelastic overlap function $h_{inel}(s,b)$ from low to  high energies.}
\end{figure}
Maximal value of $h_{inel}$ takes place at $b=r(s)$, where $r(s)$ is the solution of the equation $f(s,b)=1/2$. Due to such explicit peripheral $b$--dependence of $h_{inel}(s,b)$ any observable describing a process of multiparticle production, $A(s,\xi)$ can be expressed in the following form at $s\to\infty$ \cite{prod14}:
\begin{equation}
\label{ask}
A(s,\xi) \simeq A(s,b,\xi)|_{b=r(s)}.
\end{equation}
Comparing the existing low energy dependence of the multiparticle production observables with the asymptotic one given by Eq. (\ref{ask}), one can  state on their energy evolution at the LHC. In particular, it has been concluded that the mean multiplicity $\langle n\rangle (s)$ and the mean transverse momentum $\langle p_T\rangle(s)$ should start to change their energy dependence at the LHC, namely a slow down of their energy increase should take place due to appearing the reflective scattering mode, which assumes domination of the geometric scattering in the central hadron collisions. 
    
Indeed,  the   recent experimental data at  $\sqrt{s}=7$ TeV \cite{cms7} have started to 
demonstrate deviation from the trends observed at lower energies.  This is related to the energy 
dependence of the average transverse momentum of the secondary particles (Fig.5).
\begin{figure}[h]
	\begin{center}
		\resizebox{8cm}{!}{\includegraphics*{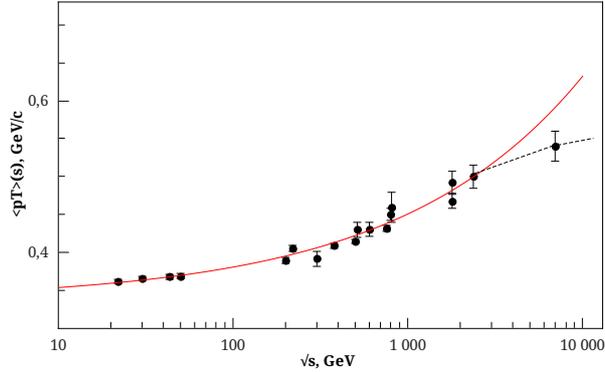}}
	\end{center}
	\caption{{Slow down of the $\langle p_T\rangle(s)$} dependence \cite{decoh}.}
\end{figure}
There is only data for multiplicity at central rapidities at the LHC, however the empirical 
relation \cite{decoh}
\begin{equation}\label{ptn}
\langle p_T\rangle(s)=a+b\langle n\rangle (s). 
\end{equation}
allows one to conclude on the similar slow down in the energy evolution of the mean multiplicity at the LHC.

It may be asked at the moment what can be said about energy dependence of the slope parameter $B(s)$.  Contrary to the case of absorptive scattering, we do not expect any changes in $B(s)$--dependence if the reflective scattering mode starts to develop.  At the asymptotic energies, i.e. in the limit $s\to\infty$, the following inequality\cite{webb} can be applied:
\begin{equation}
\sqrt{B(s)}\geq \frac{1}{2\langle p_T\rangle(s)\langle n \rangle(s)}.
\end{equation}
and one can speculate on the effect of slow down of mean multiplicity and mean transverse momentum of the secondary particles in multiparticle production processes on the slope of the diffraction cone in elastic scattering.

\section{Deep-elastic scattering at the LHC}
Transition to the reflective scattering mode can be related to increasing central density of the colliding protons with the  energy increase. One can imagine that beyond some critical value of this density (which correspond to the black-disk limit) the colliding protons start acting  like the billiard balls under collision. This effect seems to be similar to  reflection of the incoming wave of light by a metal (it changes  phase of incoming wave by $180^0$ due to presence of free electrons in a metal).  Then increasing reflection ability   is correlated with  a decreasing absorption due to probability conservation which results in the unitarity relation. 
The reflective scattering  mode is  characterized by the inequalities  $1/2  < f(s,b) <1$ and $0  > S(s,b) > -1$,  which are allowed by the unitarity  \cite{bd1,bd2}.

 Exceeding the  black-disk limit is expected to occur first in central hadron collisions, i.e in the vicinity of $b=0$. 
The amplitude $f(s,b)$  at small values of $b$  is mostly sensitive to the $t$--dependence of the scattering amplitude $F(s,t)$ in the region of large values of $-t$ (known as deep--elastic scattering \cite{islam}).   In \cite{deepel} it has been shown that  saturation of the unitarity limit results in  the relation
\begin{equation}\label{asmpt}
(d\sigma^{rfl}_{deepel}/dt)/(d\sigma^{abs}_{deepel}/dt)\simeq 4.
\end{equation}
Since at the LHC energies  the difference of the absorptive and reflective scattering modes is not too significant (positive deviation from the black-disk limit is small), the above ratio has an approximate value
\begin{equation}\label{asmpt1}
(d\sigma^{rfl}_{deepel}/dt)/(d\sigma^{abs}_{deepel}/dt)\sim 1+2\alpha(s,b=0).
\end{equation}
The models based on absorptive approach do not reproduce crossing of the black-disk limit, in particular, the standard eikonal models, do not provide a good description of the LHC data in the deep--elastic scattering region (cf. e.g. \cite{b2,godiz,drufn}). 
With  a rather poor  description of the differential cross-section in this region of $-t$ the existing crossing of the black-disk limit can be  missed.  This fact emphasizes again an importance of consideration of the {\it unintegrated} quantities. Contrary to absorptive models,  the rational  form of unitarization of the scattering amplitude  is in a better agreement with large--$t$ data (cf. Fig. 3). It is also valid for
the Donnachie-Landshoff  model \cite{dln} and the reason is that it is an another  model where the black--disk limit is crossed.

\section{Conclusion}
The two diskussed scattering modes---saturation of the black disk limit (absorptive mode) or saturation of the unitarity limit (reflective mode)---corresponds to the different mechanisms of the total cross--section growth at the LHC energies. It is important that the black--disk limit is commonly considered to be reached at the LHC. Saturation of the black disk limit correspond then to the  total cross--section increase due to expanding impact parameter range only, while the  total cross-section growth can, in fact, be due to both factors --- expansion of the impact parameter range combined with an increase of the elastic scattering amplitude $f(s,b)$ with energy at fixed impact parameters values.   Therefore,  measurements of the elastic scattering at small and large values of the transferred momentum at different LHC energies and obtaining the related knowledge  energy evolution of the slope of the diffraction cone  at these energies  would help to determine which  scattering mode ---absorptive or reflective --- is to be expected  asymptotically.

\end{document}